\documentclass[conference]{IEEEtran}
\IEEEoverridecommandlockouts
\usepackage{cellspace}

\usepackage{cite}
\usepackage{array}
\usepackage{amsmath,amssymb,amsfonts}
\usepackage{algorithmic}
\usepackage{graphicx}
\usepackage{textcomp}
\usepackage{xcolor}
\def\BibTeX{{\rm B\kern-.05em{\sc i\kern-.025em b}\kern-.08em
    T\kern-.1667em\lower.7ex\hbox{E}\kern-.125emX}}
\begin{document}

\title{EEG-estimated functional connectivity, and not behavior, differentiates Parkinson's patients from health controls during the Simon conflict task\\
\thanks{This work was supported by a Vannevar Bush Faculty Fellowship from the US Department of Defense (N00014-20-1-2027) and a Center of Excellence grant from the Air Force Office of Scientific Research (FA9550-22-1-0337). All authors are affiliated with the Department of Biomedical Engineering. Dr. Sajda is also affiliated with the Department of Electrical Engineering, Radiology, and Data Science Institute.}
}

\author{\IEEEauthorblockN{1\textsuperscript{st} Xiaoxiao Sun}
\IEEEauthorblockA{
\textit{Columbia University}\\
New York, U.S.A \\
xiaoxiao.sun@columbia.edu\vspace{-0.8cm}}
\and
\IEEEauthorblockN{2\textsuperscript{nd} Chongkun Zhao}
\IEEEauthorblockA{
\textit{Columbia University}\\
New York, U.S.A \\
cz2715@columbia.edu\vspace{-0.8cm}}
\and
\IEEEauthorblockN{3\textsuperscript{rd} Sharath Koorathota}
\IEEEauthorblockA{
\textit{Columbia University}\\
New York, U.S.A \\
sk4172@columbia.edu\vspace{-0.8cm}}
\and
\IEEEauthorblockN{4\textsuperscript{th} Paul Sajda}
\IEEEauthorblockA{
\textit{Columbia University}\\
New York, U.S.A \\
psajda@columbia.edu\vspace{-0.8cm}}
}
\maketitle
\begingroup\renewcommand\thefootnote{\textsection}

\endgroup

\begin{abstract}
Neural biomarkers that can classify or predict disease are of broad interest to the neurological and psychiatric communities. Such biomarkers can be informative of disease state or treatment efficacy,  even before there are changes in symptoms and/or behavior. This work investigates EEG-estimated functional connectivity (FC) as a Parkinson's Disease (PD) biomarker. Specifically, we investigate FC mediated via neural oscillations and consider such activity during the Simons conflict task. This task yields sensory-motor conflict, and one might expect differences in behavior between PD patients and healthy controls (HCs). In addition to considering spatially focused approaches, such as FC, as a biomarker, we also consider temporal biomarkers, which are more sensitive to ongoing changes in neural activity. We find that FC, estimated from delta (1-4Hz) and theta (4-7Hz) oscillations, yields spatial FC patterns significantly better at distinguishing PD from HC than temporal features or behavior. This study reinforces that FC in spectral bands is informative of differences in brain-wide processes and can serve as a biomarker distinguishing normal brain function from that seen in disease.


\end{abstract}


\section{Introduction}

With the advent of high-density neural recording tools (e.g., Neuropixels), there is an increasing interest in  investigating neural activity in localized brain regions~\cite{kahana2006cognitive}. An alternative approach is to consider brain-wide activity, albeit at a coarser temporal scale (e.g., not measuring spiking activity), since the technology does not exist for such spatial coverage at the resolution of neuronal spikes, particularly for humans. Many brain-wide analyses have instead focused on lower temporal resolution neural oscillations (1-40Hz). Such oscillations are believed to link neural activity across disparate regions, potentially binding activity in a way that yields our integrated and coherent perception of the world~\cite{ward2003synchronous}. 

Recent studies have shown brain oscillations drive functional connectivity (FC) across cortical and subcortical networks~\cite{fox2012efficacy,bonnefond2017communication}. Moreover, the specific spatiotemporal properties of such FC appear to correlate with normal cognitive, emotional, and perceptual functions as well as neuropathologies~\cite{smith2008spatial,northoff2020spatiotemporal}. For instance, investigating differences in FC between disorders such as schizophrenia and bipolar disorder has provided insights into possible disease mechanisms~\cite{birur2017brain}. This growing body of research has highlighted the potential utility of FC as a biomarker for identifying cognitive function and diagnosing brain disorders.  
 
In this study, we investigate EEG-estimated FC as a potential biomarker for Parkinson's disease (PD) when both PD patients and healthy controls (HCs) are engaged in the Simon conflict task. Using different machine learning (ML) models, we systematically assess the relative importance of temporal and spatial-focused features in classifying PD vs. HC. Our findings show that spatial features represented by EEG-estimated FC are extremely good at classifying PD vs. HC, while models focusing on temporal features perform substantially worse. Notably, neither spatial-FC features nor temporal features are particularly good at classifying trial conditions (i.e., congruent vs incongruent trials). This suggests that FC is a potential spatial biomarker for PD that is not simply an artifact of brain signals that differentiate task performance or task condition. Based on this finding, we also investigated the significance of functional connectivity patterns using a statistical approach -- the surrogate test. The results reflect a similar interpretation, evidenced by the higher number of electrode pairs that remain relevant after applying a threshold for group comparisons, as opposed to a reduced number of pairs for trial condition comparisons. 



\section{Related Work}

\subsection{Brain Oscillations and Cognitive Coordination}

Neuronal oscillations are believed to play various roles in brain function~\cite{bonnefond2017communication}. Because neuronal oscillations can change the dynamic interactions between brain regions, they have been proposed as a mechanistic gate for routing information on a fast time scale~\cite{varela2001brainweb}.  Oscillations in different frequency bands are believed to be essential for flexible and coordinated brain function, with recurrent connections between excitatory and inhibitory cell populations producing resonant oscillation frequencies that depend on the decay time constants in the neural populations~\cite{kang2010lfp}. 

\subsection{Simon Conflict Task}
The Simon conflict task is a widely employed cognitive paradigm designed to investigate the influence of conflicting information on response inhibition and decision-making processes. The task engages participants in a visual-spatial conflict by presenting stimuli in a way that creates a misalignment between the spatial location of the stimulus and the required response~\cite{simon1963}. It has been shown valuable for probing the mechanisms underlying cognitive control, attentional processing, and the brain's ability to resolve conflicting information~\cite{singh2023evoked}. The task requires flexible and coordinated processing that is typically characterized by the term ``cognitive control".

\subsection{Functional Connectivity}
Oscillatory neuronal activity may provide a mechanism for dynamic network coordination and be the basis for what is typically called functional connectivity (FC). FC is believed to exist, and even change, while an individual is at rest or engaged in a task. FC is thought to be crucial for governing information processing and functional execution across diverse brain areas~\cite{bonnefond2017communication,pantazatos2023timing}. Population differences in FC have also been investigated as being potential biomarkers of neurological disease and psychiatric illness ~\cite{du2018classification,birur2017brain,tian2021machine,bratic2018machine}. In many cases, machine learning is used to characterize features of FC and classify individuals or populations. 


\subsection{Machine Learning Models}
\paragraph{Random Forest (RF) Models}
Random Forest (RF) is a widely employed machine learning method for classification problems~\cite{breiman2001random}. Based on the ensemble learning paradigm, RF leverages the power of decision trees by constructing an ensemble of them and subsequently combining their predictions~\cite{belgiu2016random}. By aggregating the outcomes of multiple decision trees through a voting mechanism, RF models offer robustness against noise and outliers, making it a versatile tool in various domains. 

\paragraph{Long Short-Term Memory (LSTM) Networks}
A Long Short-Term Memory (LSTM) network is a type of recurrent neural network (RNN) architecture that has gained prominence in deep learning due to its ability to capture and model long-range dependencies in sequential data~\cite{yu2019review}. Unlike traditional RNNs, LSTM utilizes specialized gating mechanisms, including input, forget, and output gates, which enable them to selectively store and update information over extended sequences, thus mitigating the vanishing gradient problem~\cite{hochreiter1997long}. Owing to their proficiency in handling sequential data of varying lengths and complexities, LSTMs have found widespread applications across various fields, including natural language processing, speech recognition, time series forecasting, and neural data analysis.

\paragraph{Convolution Neural Networks (CNN)}
Convolution Neural Networks (CNNs) have revolutionized the field of computer vision and image analysis due to  their exceptional ability to extract hierarchical features from visual data~\cite{lecun1998gradient}. The architecture's success can be attributed to its ability to learn discriminative features from data through training. These neural networks, inspired by the human visual system, consist of multiple layers of learnable filters that perform convolution operations on input images, enabling them to capture local and global patterns~\cite{sharma2020performance}. 
\vspace{-0.15cm}
\section{Methodology}

\subsection{Experimental Dataset}\label{subsec:dataset}
\begin{figure}[htbp]
\centerline{\includegraphics[width = 1\linewidth]{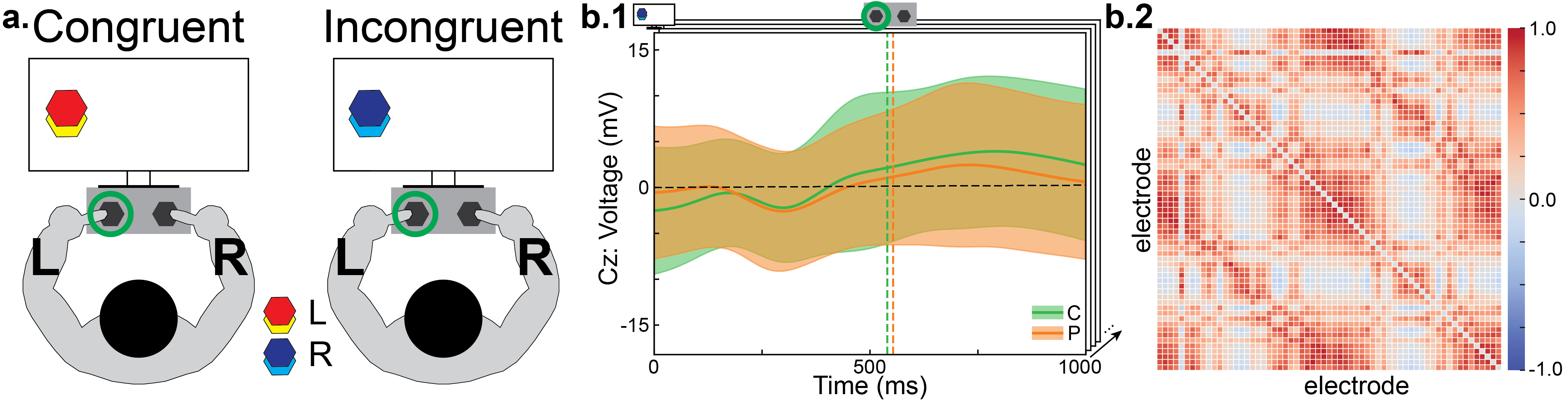}}
\vspace{-0.3cm}
\caption{Overview of the experimental setup and the extracted temporal/spatial features. \textbf{a.} 
Illustration of the Simon conflict task (see text for details). \textbf{b.1} Epoched 1-second temporal feature after the stimulus onset, specifically the voltage at electrode Cz within the delta band for incongruent trials. The average response time (RT) is $549.6\pm113.1$ ms for HCs and $563.2\pm156.2$ ms for PD patients in the incongruent trials (highlighted with the vertical dashed line separately, see Fig.~\ref{fig:binary}\textbf{a.1} and \textbf{a.2} for more details about RT). \textbf{b.2} A corresponding spatial feature from the same time window. The difference in functional connectivity (represented by Pearson Correlation) between HC and PD patients is plotted as an example.
}
\label{fig:experiment}
\vspace{-0.7cm}
\end{figure}

We investigated functional connectivity within the context of the Simon conflict task using an open-source EEG dataset~\cite{ds004580:1.0.0}. In the task, a stimulus is presented on either the left or right side of the screen and participants are instructed to press the left key when the stimulus is yellow/red and the right key when it is cyan/blue~\cite{singh2023evoked}. A congruent trial is when the stimuli match the side of the screen where the response hand is located, while an incongruent trial is when the stimuli appear on the opposite side of the screen from the response hand (i.e., spatially congruent/incongruent, see Fig.~\ref{fig:experiment}\textbf{a}).

\vspace{-0.2cm}

\begin{table}[htbp]
\caption{Experiment Data Summary}
\vspace{-0.4cm}
\begin{center}
\begin{tabular}{|c|c|c|c|c|} 
\hline
\textbf{\# of Trials} &  \textbf{Patient (49)} & \textbf{Control (26)} & \textbf{Total} & \textbf{Average}$^{\mathrm{a}}$\\
\hline 
\textbf{Congruent} & 7422 &  4456 & 11878 & 158$\pm$27 \\
\hline
\textbf{Incongruent} & 6655 & 4235 & 10890 & 145$\pm$34\\
\hline
\textbf{Total} & 14077 & 8691 & 22768 & - \\
\hline
\end{tabular}\\
\vspace{0.1cm}
$^{\mathrm{a}}$Average within each subject by trial condition. 
\label{tab:exp}
\vspace{-0.5cm}
\end{center}
\end{table}

Our analysis includes a subset of 75 subjects from the open-source dataset, encompassing 49 patients with PD and 26 HCs (see experiment summary in Table~\ref{tab:exp}). EEG was recorded using a 64-channel BrainVision system. The EEG was high-pass filtered at 0.1Hz and down-sampled to 500Hz. Electrode Pz was used as the reference, and electrode Fpz was used as the ground. EEG activity at electrode FT9 was excluded since a specialized cap without FT9 was employed, resulting in 63 channels for further analysis (see TABLE~\ref{tab:ML}). Additional details about subject recruitment and data recording are available in their latest paper~\cite{singh2023evoked}. 

\subsection{Temporal Biomarkers: Band-passed electrode time series (BPT)}
Since low-frequency oscillations are  modulated by conflict, attention, and timing and are also known to be specifically impaired in PD~\cite{cavanagh2014frontal,singh2021timing}, we focus our analysis of temporal biomarkers in delta (1 to 4Hz) and theta (4 to 7Hz) bands of the EEG. Bidirectional filtering was applied to the raw EEG data using a bandpass filter (Matlab2023a) to attenuate frequencies outside the delta and theta bands. Event-related potentials (ERPs) were extracted with respect to the onset of the stimuli (marked as 0 seconds), spanning from -0.5 seconds before to 1.5 seconds after the stimulus. Baseline removal was done based on the entire epoch. To align the brain activity with the behavioral response, we exclusively considered the 1-second interval following the stimulus onset (see Fig.~\ref{fig:experiment}\textbf{b.1} and \textbf{b.2}), as the maximum response time (RT) typically falls within this time frame (see Fig.~\ref{fig:binary}\textbf{a.1} and \textbf{a.2}). This extracted 1-second band-passed time series (BPT) of all electrodes were then used as temporal feature inputs for ML models (see TABLE~\ref{tab:ML}).  

\subsection{Spatial Biomarkers: Functional Connectivity (FC)}
 FC can be quantified using  multiple different metrics, each possessing its own strengths and limitations~\cite{bastos2016tutorial}. Here, we employed three non-directed methods for FC measurement, encompassing amplitude-based, phase-based, and amplitude-phase-mixed approaches.

\subsubsection{Pearson Correlation (PC)}
The simplest metric for non-directed interactions is the Pearson correlation coefficient, which measures the linear relationship between two random variables~\cite{arbabshirani2014impact}. PC focuses on the dynamic connection between regions represented by their voltage (i.e., amplitude) changes.  
\begin{equation}\label{eq:pc}
    \text{PC}(x,y) = \frac{\sum^n_{i=1}(x(\omega,i)-\bar{x})(y(\omega,i)-\bar{y})}{\sqrt{\sum^n_{i=1}(x(\omega,i)-\bar{x})^2\sum^n_{i=1}(y(\omega,i)-\bar{y})^2}}
\end{equation}
where $x(\omega,i)$ ($y(\omega,i)$) are the values of $x$ ($y$) in a sample $i$ of target frequency band $\omega$ (i.e., theta, alpha, etc.); $n$ is the number of samples; $\bar{x}$ ($\bar{y}$) is the mean of $x$ ($y$). 

\subsubsection{Phase Locking Value (PLV)}
An alternative metric to measure FC is to evaluate the synchronization between cortical areas (e.g., between electrodes from EEG recordings). PLV estimates how the relative phase between signals is distributed over the unit circle~\cite{niso2013hermes}. PLV is a scalar value that ranges from $[0, 1]$, where 1 indicates the strongest phase synchronization between regions/electrodes~\cite{sun2023increased}. 
\begin{equation}\label{eq:plv}
\begin{split}
    \text{PLV}(x,y) &= \frac{|\frac{1}{n}\sum^n_{i=1}1_x(\omega,i)1_y(\omega,i)e^{i(\phi_x(\omega,i) - \phi_y(\omega,i))}|}{\sqrt{(\frac{1}{n}\sum^n_{i=1} 1_x^2(\omega,i))(\frac{1}{n}\sum^n_{i=1} 1_y^2(\omega,i)})}
\end{split}
\end{equation}
where $\phi_x(\omega,i)$ ($\phi_y(\omega,i)$) is the phase of $x$ ($y$) obtained by the Hilbert transform in a sample $i$ of targeted frequency band $\omega$; $n$ is the number of samples.

\subsubsection{Phase Coherence (COH)}
A measurement similar to PLV that represents FC with phase synchronization is COH. It is basically an extension of PLV. Instead of only considering the phase after the Hilbert transform, COH is computed from both the amplitude and the phase~\cite{bastos2016tutorial}.

\begin{equation}\label{eq:coh}
    \text{COH}(x,y) = \frac{|\frac{1}{n}\sum^n_{i=1}A_x(\omega,i)A_y(\omega,i)e^{i(\phi_x(\omega,i) - \phi_y(\omega,i))}|}{\sqrt{(\frac{1}{n}\sum^n_{i=1} A_x^2(\omega,i))(\frac{1}{n}\sum^n_{i=1}A_y^2(\omega,i)})}
\end{equation}
where $A_x(\omega,i)$ ($A_y(\omega,i)$) is the amplitude and $\phi_x(\omega,i)$ ($\phi_y(\omega,i)$) is the phase of $x$ ($y$) obtained by Hilbert transform in a sample $i$ of targeted frequency band $\omega$; $n$ is the number of samples.

\subsection{Comparing Machine Learning Models}
We used the RF model as a benchmark model for comparing classification performance and assessing the relative effectiveness of temporal (i.e., voltage changes of ERPs) and spatial (i.e., FC) features. LSTMs were used to analyze temporal features, and CNNs  used for spatial features, given that their designs matched these dimensions. We also compared the spatial feature weights extracted from different models to gain deeper insight into the importance of FC features across different ML models. By employing this diverse set of ML architectures, our aim was to  comprehensively explore the features embedded within the neural data (see Table~\ref{tab:ML}). The dataset (see Table~\ref{tab:exp}) was divided into three subsets: 64\% for training, 16\% for validation, and the remaining 20\% for testing. We considered the imbalanced dataset distribution to ensure robust and accurate results and applied a 10-fold cross-validation during training.

\vspace{-0.5cm}
\begin{table}[htbp]
\caption{Model Architectures Summary}
\vspace{-0.4cm}
\begin{center}
\begin{tabular}{|c|c|c|c|} 
\hline
\textbf{Model} &  \textbf{Input Shape} & \textbf{Layers} & \textbf{Para Size}\\
\hline 
RF$^{\mathrm{a}}$ & (31500, ) &  - & -  \\
\hline
RF$^{\mathrm{b}}$ & (1953, ) & - & - \\
\hline
LSTM1$^{\mathrm{c}}$ & (63, 500) & (256, 256)-256 & 3.51MB \\
\hline
LSTM2 & (63, 500) &(256, 128)-256 & 2.13MB \\
\hline
LSTM3 & (63, 500) &(128, 128)-256 & 1.01MB \\
\hline
LSTM4 & (63, 500) &(128, 64)-256 & 647.00KB \\
\hline
LSTM5 & (63, 500) &(64, 64)-256 & 327.00KB \\
\hline
LSTM6 & (63, 500) &(64, 32)-256 & 214.50KB \\
\hline
CNN1$^{\mathrm{d}}$ & (63, 63) &(32)-256 & 26.29MB \\
\hline
CNN2 & (63, 63) &(16)-256 & 13.15MB \\
\hline
CNN3 & (63, 63) &(16)-128 & 6.58MB \\
\hline
CNN4 & (63, 63) &(16)-64 & 3.29MB \\
\hline
CNN5 & (63, 63) &(16)-32 & 1.65MB  \\
\hline
CNN6 & (63, 63) &(8)-32 & 842.69KB  \\
\hline
\end{tabular}\\
\vspace{0.1cm}
$^{\mathrm{a}}$Input shape is transformed into a flattened representation (31500=63$\times$500). $^{\mathrm{b}}$Input shape is transformed into a flattened representation, and repeated values have been removed (1953=63$\times$62/2). $^{\mathrm{c}}$The number within parentheses represents the dimensionality of the LSTM layer, and the value following the hyphen(-) indicates the size of the dense layer. $^{\mathrm{d}}$The number within parentheses represents the dimensionality of the CNN layer, and the value following the hyphen indicates the size of the dense layer.
\label{tab:ML}
\vspace{-0.5cm}
\end{center}
\end{table}

\begin{figure*}[htbp]
    \centering
    \includegraphics[width=1\linewidth]{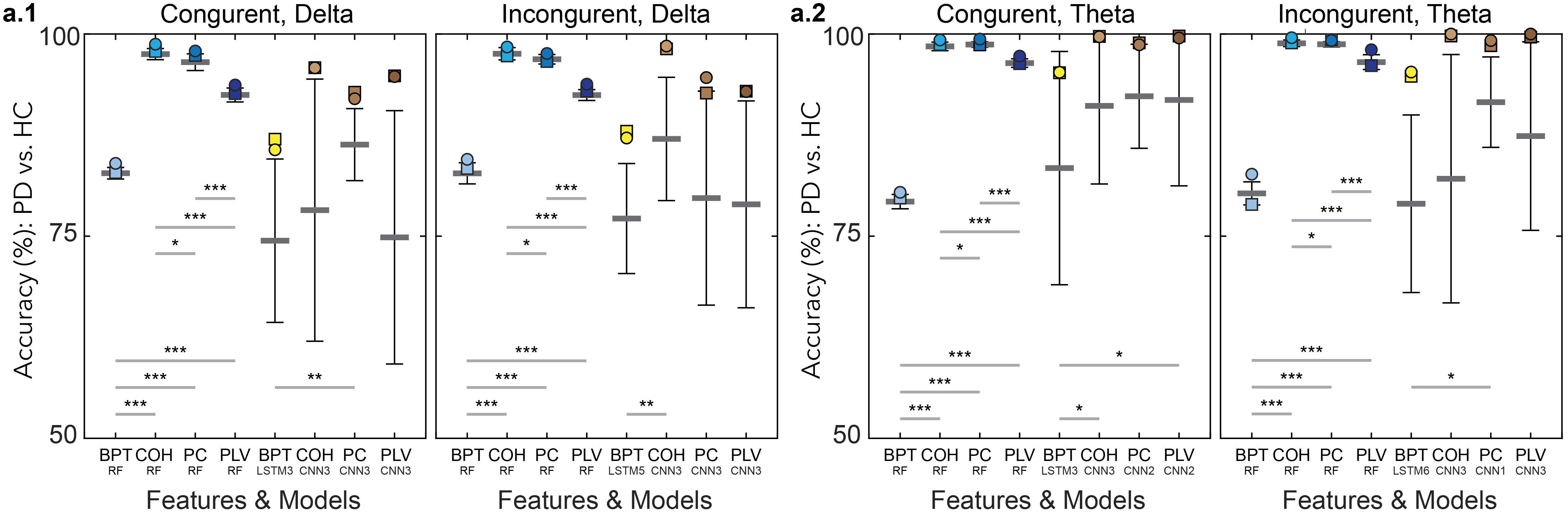}
    \vspace{-0.7cm}
    \caption{Prediction accuracy (HC vs. PD) from ML models. \textbf{a.1} Model performance with features in the delta band (1-4Hz). Among the LSTM and CNN models trained and evaluated (see Table~\ref{tab:ML}), we show the  model having the highest test accuracy (highlighted with square markers). The bar plots are generated using 10-fold cross-validation results (mean $\pm$ standard deviation). Each model's highest validation accuracy is highlighted with a circle marker. We used the Mann-Whitney-Wilcoxon test given the small sample sizes (n=10)~\cite{fay2010wilcoxon}. The significance level of the test result (multi-comparison un-corrected) is indicated by a black asterisk, where *** indicates significance under a 99.9\% confidence interval, ** indicates significance under a 99\% confidence interval, and * indicates significance under a 95\% confidence interval.  
    \textbf{a.2} is similar to \textbf{a.1}, except that models are evaluated in the theta band (4-7Hz).}
    \label{fig:acc}
    \vspace{-0.6cm}
\end{figure*}

For all the FC approaches, we use the BPT data to compute the FCs.  Thus, both the BPT and FC use the same time window of EEG and frequency bands, however, BPT maintains the time-series information while FC collapses it into a spatial connectivity measure across the brain.

\subsection{Surrogate Tests}
\label{sec:surrogate}
Since phase-related functional connectivity metrics such as PLV and COH are usually highly sensitive and can yield spurious results (e.g., two narrow-band Gaussian white noise processes can generate a moderate phase-related index), we employed the surrogate test to detect the statistically significant patterns in FC represented by PLV and COH~\cite{xu2006spurious}. The essence of the surrogate test is to test statistical significance from an artificially generated null distribution of surrogate signals that simulate the Fourier Power Spectrum of the original signal. To generate this null distribution, we use an unwindowed Fourier transform algorithm~\cite{theiler1992testing}, which randomizes the phases by multiplying each instantaneous phase  by $e^{i\phi}$, where $\phi$ is independently chosen from a uniform distribution $U([0, 2\pi])$.

 We simulated 100 surrogate signals of the original signal for each trial and computed the corresponding surrogate FC in PLV(COH). The statistical significance of each PLV(COH) measurement was assessed by comparing it to the distribution of surrogate PLVs (COHs) ($p < 0.05$). Repeating this method for all electrode pairs for one trial returned a subset of statistically significant electrode pairs, represented by a feature matrix where 1 indicates a significant pair and 0 indicates an insignificant pair.

\section{Results}

\subsection{Model Prediction Efficiency and Accuracy}
We investigated two binary classification tasks: group differentiation (PD vs. HC), which is our primary goal, and trial condition differentiation (congruent vs. incongruent). The reason for the latter is that we wanted to check that any potential biomarker of group differentiation was not simply due to detecting differences in how the two populations (PD and HC) did the task. We found for all models and both biomarker types (BPT and FC), there was no ability to discriminate between the trial conditions (mean accuracy 51\%, $p = 0.564$). We did, however, find significant discrimination of group (PD vs HC) using the biomarkers, across all ML model types (see Table~\ref{tab:ML}).  We saw this discrimination both in the congruent trial and incongruent trial data, further indicating that the biomarkers were not artifacts of task condition differences but rather intrinsic disease states.

The spatial-focused features (i.e., FC) consistently outperformed temporal features (i.e., BPT) in both the delta and theta bands. Within the delta band, the amplitude-phase-mixed FC metric, i.e., COH, showed the highest validation and test accuracy using the RF model among congruent trials and the CNN model among incongruent trials (see Fig.~\ref{fig:acc}\textbf{a.1}). Amplitude-based prediction using the theta band performed better than the delta band ($87\%$ in delta vs $95\%$ in theta with the LSTM).  The amplitude-based FC metric, i.e., PC, results in the highest validation and test accuracy using the RF model for the data from the congruent trials and the incongruent trials (see Fig.~\ref{fig:acc}\textbf{a.2}). For CNN models, COH-derived FC usually generates the best performance, though when considered statistically, there were no significant performance differences  between the different FC metrics. Although similar test accuracies are obtained in both RF and CNN models given spatial features as input, the RF model showed more robust performance across validation folds, reflected in the smaller variance in validation accuracy. 

\begin{figure}[htbp]
    \centering
    \includegraphics[width=1\linewidth]{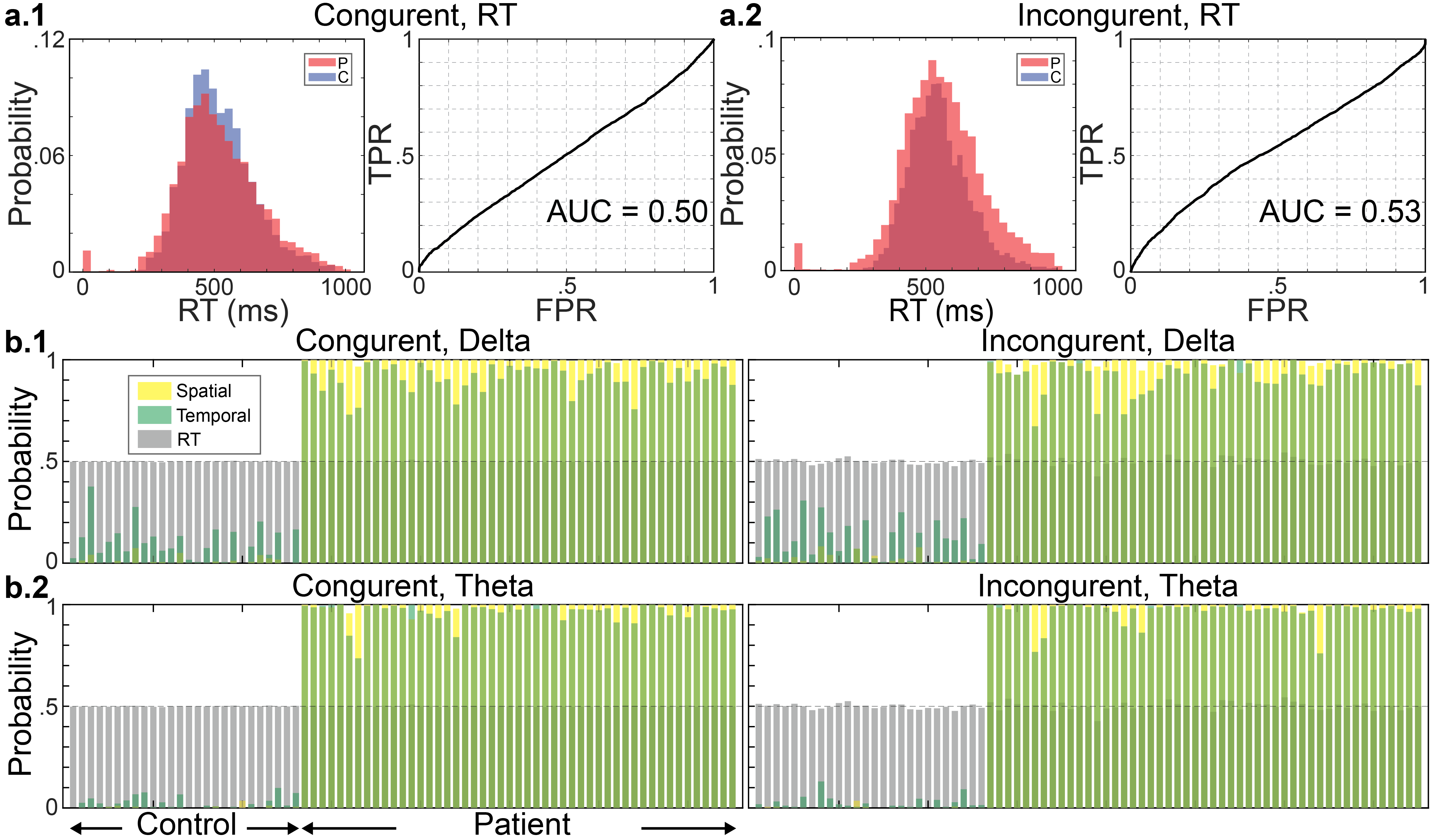}
    \vspace{-0.7cm}
    \caption{Classification probability given different biomarkers. \textbf{a.1} Histogram of RT and receiver operating characteristic (ROC) curve of HC vs PD classification based on reaction time (RT). \textbf{a.2} is similar to \textbf{a.1}, but for the incongruent trials. \textbf{b.1} Subject-level probabilities of classifying HC vs PD. Both spatial FC (yellow) and temporal BPT (green) biomarkers outperform RT (gray), with 26 controls having probabilities less than 0.5 and 49 patients having probabilities greater than 0.5. The spatial FC biomarkers are clearly more accurate than the temporal BIT biomarkers across the population. }
    \label{fig:binary}
    \vspace{-0.4cm}
\end{figure}

We used the best model (i.e., has the highest test accuracy) obtained from trial-level prediction of PD vs HC  to predict the group condition at the subject-level. We did this by computing the fraction of trials that the subject was characterized as PD vs HC, and we call this the subject-level probability. We also compared the results to just using behavior in the task (classifying by RT).  Fig.~\ref{fig:binary}\textbf{b.1} shows the probabilities of one subject being classified as a PD patient with FC calculated in the delta band. Notably, both temporal and spatial features effectively distinguish between different group conditions. However, the models utilizing spatial features have better accuracy, with the subject-level probability scores being  closer to 0 for HC subjects and closer to 1 for PD patients. This distinction underscores the efficacy of spatial features in effectively differentiating between groups on an individual level. Analyses within the theta band yielded similar results (see Fig.~\ref{fig:binary}\textbf{b.2}).

\subsection{Evaluation of Spatial Biomarker Patterns using Surrogate Tests}
\begin{figure}[htbp]
    \centering
    \includegraphics[width=1\linewidth]{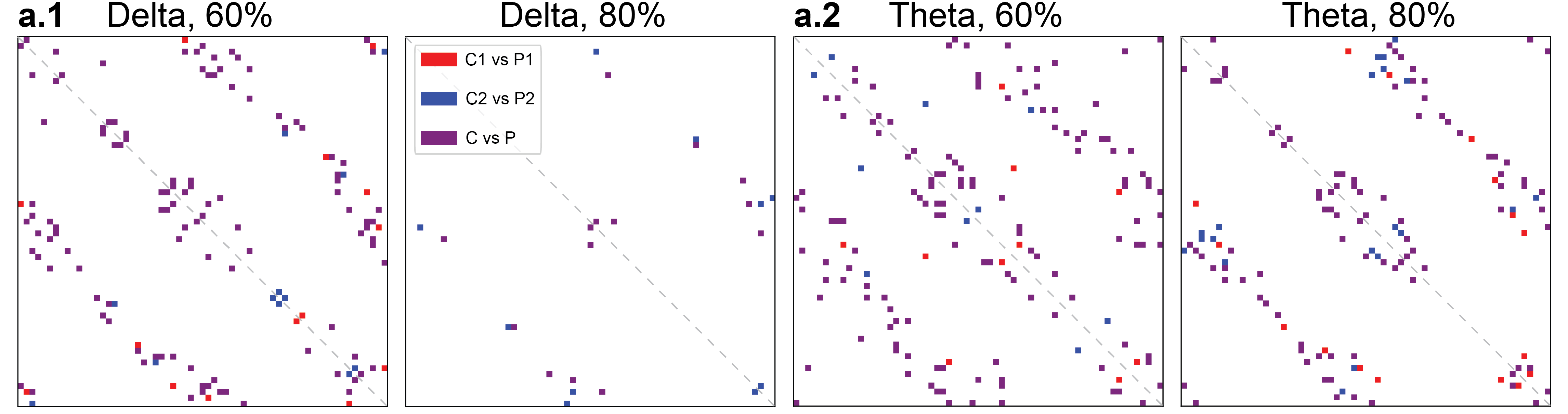}
    \vspace{-0.7cm}
    \caption{The surrogate test identified significantly ($p<0.05$) different functional connectivity (FC) patterns under different thresholds (60\% and 80\%). \textbf{a.1} Different FC patterns between groups within the delta band. Each color represents a comparison within one trial condition (red: congruent, labeled as 1 in the legend, e.g., C1 means HC-congruent; blue: incongruent, labeled as 2 in the legend, purple: shared in both conditions). \textbf{a.2} follows the layout of \textbf{a.1}, except that it is computed based on the theta band. }
    \label{fig:surr_pattern}
    
\end{figure}

We performed the surrogate test analysis to compare the spatial FC patterns between HC and PD computed for different trial conditions (congruent or incongruent). We first computed a significance mask across the FC dimensions using a $p < 0.05$ criterion under the null hypothesis created by the surrogate data. We applied this mask to all trials for each condition and then computed the percentage of trials that passed this test against the null (see Section~\ref{sec:surrogate}). We considered two thresholds, 60\% and 80\%. For example, a 60\%  threshold in Fig.~\ref{fig:surr_pattern} represents FC patterns that pass the surrogate test for more than 60\% of all trials. We compared the FC network patterns between groups (HC vs. PD). We see that most of the dimensions of the FC biomarkers are common across the congruent and incongruent trials (purple color Fig.~\ref{fig:surr_pattern}). This is consistent with our previous findings which showed that the ML classifiers could not distinguish trial type using the FC biomarker --i.e. there is no difference between the congruent or incongruent trial types. In addition, we see that the theta band derived FC has more robust FC dimensions relative to those derived from the delta band-i.e. more dimensions survive the surrogate test at the 80\% threshold in theta than for delta. 


\vspace{-0.1cm}

\section{Discussion and Conclusion}
This work investigated spatial and temporal biomarkers of Parkinson's disease (PD) derived from EEG data recording while participants performed the Simon conflict task. We found that spatial biomarkers, specifically those constructed to represent functional connectivity, were superior to temporal biomarkers in classifying PD vs. HC. Noteworthy was that the FC and BPT temporal features could not be used to classify trial conditions (congruent vs. incongruent), suggesting that the underlying biomarkers were not task-specific but related to fundamental differences in brain function across the groups and independent of the task. We did not have data to test whether PD vs. HC classification was possible with these biomarkers in resting state EEG, or during another task, so it remains to be seen if the performance of the task is critical for inducing brain states that differentiate the two populations. Given that our analysis was event-locked (to the stimulus presented in this case), it does suggest that a task is desirable for aligning the neural data to construct the biomarkers. 

We further explored the significance of the FC biomarkers via a statistical surrogate test with results consistent with our ML results, namely that the FC biomarkers capture spatial information that differentiates PD from HC and which is not simply a result of differences in task performance or trial condition. Our work supports the  potential diagnostic value of spatial features, specifically functional connectivity derived from EEG measures.


\bibliographystyle{ieeetr}
\bibliography{references}

\end{document}